\documentclass[aps,twocolumn,a4paper,pra,superscriptaddress,floatfix,showpacs]{revtex4}%
\usepackage{amssymb}
\usepackage{amsmath}
\usepackage{amsfonts}
\usepackage{graphicx}
\usepackage{graphics}

\def\Drud{\mathrm{Drud}}
\def\txi{{\tilde \xi}}
\def\plas{\mathrm{plas}}
\def\perf{\mathrm{perf}}
\def\cL{\mathcal{L}}
\def\cF{\mathcal{F}}
\def\cM{\mathcal{M}}
\def\cR{\mathcal{R}}
\def\cK{\mathcal{K}}
\def\P{\mathrm{P}}    %plane
\def\S{\mathrm{S}}    %sphere
\def\TE{\mathrm{TE}}
\def\TM{\mathrm{TM}}

\def\br{\mathbf{r}}
\def\bk{\mathbf{k}}

\def\bK{\mathbf{K}}

\def\hK{\mathbf{\hat K}}

\begin{document}

\title{Thermal Casimir effect for Drude metals in the plane-sphere geometry}

\author{Antoine Canaguier-Durand}
\affiliation{Laboratoire Kastler Brossel,
CNRS, ENS, Universit\'e Pierre et Marie Curie case 74,
Campus Jussieu, F-75252 Paris Cedex 05, France}

\author{Paulo A. Maia Neto}
\affiliation{Instituto de F\'{\i}sica, UFRJ,
CP 68528,   Rio de Janeiro,  RJ,  21941-909, Brazil}

\author{Astrid Lambrecht}
\author{Serge Reynaud}
\affiliation{Laboratoire Kastler Brossel,
CNRS, ENS, Universit\'e Pierre et Marie Curie case 74,
Campus Jussieu, F-75252 Paris Cedex 05, France}

\date{\today}

\begin{abstract}
We compute the Casimir interaction between a plane and a sphere, 
the configuration employed in the most precise experiments. 
The scattering formula is developed by taking a suitably chosen plane-wave 
and multipole basis and is valid for arbitrary values of the sphere radius, 
inter-plate distance, temperature and arbitrary dielectric functions 
for both sphere and plate. Our analytical and numerical results for 
metallic surfaces show a non-trivial interplay between the effects of 
curvature, temperature, finite conductivity and dissipation.
\end{abstract}


\maketitle

\section{Introduction}

Measuring the Casimir force \cite{Casimir,Lamoreaux} has been the aim
of an increasing number of experiments in the last years 
\cite{Lamoreaux97,Mohideen98,Harris00,Ederth00,Chan01,Bressi02,Decca03prl,%
ChenPRA04,DeccaPRD07,Munday07,vanZwol08,Munday09,Jourdan09,Masuda09,deMan09}.  
The comparison of these measurements with predictions from quantum electrodynamics theory
have been applied to put constraints on hypothetical new forces predicted by 
unification models \cite{OnofrioNJP06,Klimchitskaya09}.
Accurate theoretical computations, accounting for a modeling of experimental conditions,
are sorely needed for all comparisons to be reliable \cite{LambrechtNJP06,Reynaud10}.

The Casimir force is indeed very sensitive to  experimental conditions.
The effect of finite conductivity \cite{Lambrecht00} plays an essential role
in the accurate determination of the force while the contribution of thermal fluctuations
gives rise to a remarkable interplay with the latter effect 
\cite{Bostrom00,BrevikNJP06,Klim06,BrevikJPA08,Milton08R,Ingold09}.
In the calculations performed for the geometry of two parallel plates,
the Casimir force computed within the dissipative Drude model turns out
to be a factor of 2 smaller than the result obtained within the lossless plasma model.
As a consequence, the plasma theory of the Casimir effect cannot be obtained from the 
Drude one by taking the corresponding relaxation constant to zero.
The current experimental results \cite{DeccaPRD07} do not explore the region
where the calculations for lossy and lossless  models give large differences but their
precision is sufficient to favor the plasma over the Drude theory, 
in spite of the dissipative nature of the metallic plates used in the experiments.

Now, the most precise experiments are performed with a spherical metallic surface 
in front of a plane surface. The force is usually derived from the force evaluated 
in the parallel-plate geometry with subsequent averaging over the local separation distances.
This Proximity Force Approximation (PFA) \cite{Derjaguin68} is expected to provide an accurate 
description in the limit of large sphere radii (see \cite{Schaden00, Jaffe04} for derivations 
with perfect mirrors at zero temperature).
Even if spheres used in the experiments are much larger than the typical distance between them,
it remains necessary to master the beyond-PFA  geometry correction \cite{Krause07} 
even for large spheres in order to match the experimental  accuracy level.

There is no reason why the thermal, finite-conductivity and beyond-PFA corrections could be 
expected to be independent. Therefore, an accurate description of the experimental conditions 
has to take these effects into account simultaneously within a single theoretical model.
In this paper, we develop the beyond-PFA scattering approach \cite{LambrechtNJP06} 
for the plane-sphere geometry at finite temperature, with material properties described 
by either the perfect reflector, plasma or Drude models.  
We show that the interplay between temperature and material properties is drastically 
affected when the parallel-plate configuration is replaced by the plane-sphere geometry.
The results obtained from the Drude and plasma models are generally closer to each other than 
in the parallel-plate geometry. In particular, the factor of 2 between the two models 
reached at the limit of long distances between parallel plates is reduced to 3/2 for a plate 
and a sphere, and even less if small spheres are considered. 
Finally, we find that PFA underestimates the thermal contribution to the Casimir force for 
the Drude model at short distances, whereas it overestimates it at all distances for the 
perfect reflector and plasma models.

We start from the scattering theory of Casimir interaction \cite{LambrechtNJP06}, 
which allows one to consider non-trivial geometries at finite temperatures together 
with a realistic description of the material properties. 
The Casimir free energy is written in terms of general reflection operators, 
which describe non-specular diffraction by the material surfaces.
The resulting formula provides a compact way of taking into account the
multiple scatterings between the interacting bodies \cite{Balian}.
In the particular case of parallel plane surfaces,  the reflection operators are 
diagonal in the plane-wave basis, so that the Casimir free energy is given 
in terms of specular reflection coefficients \cite{JaekelReynaud91}.
In the plane-sphere geometry, scattering on the sphere is easily made explicit
in the multipole spherical-waves basis \cite{Emig08}
(see also \cite{Noguez04} for a treatment of non-retarded interactions).
It is also essential to combine the multipole basis with the plane wave basis \cite{Maia08}
well adapted to the treatment of non-ideal reflection by metals.
By judiciously employing the two bases at appropriate steps of the derivation, 
the scattering formalism thus allows to analyse the interplay between
geometry and finite conductivity at zero \cite{Canaguier09}
and non-zero temperature \cite{Canaguier10}.

Let us note that several papers
have been devoted in the recent years to the study of the Casimir effect
in non-planar geometries.
Besides the already cited papers, let us refer to the following applications 
of the scattering approach 
\cite{Bulgac06,Bordag06,Kenneth06,Emig07,Milton07,Bordag09A,Bordag09B,Rahi09} 
or of alternative methods 
\cite{Gies03,Gies06,JaffePRA05,Emig06,Dalvit06,Mazzitelli06,Rodriguez07,Weber10}.
The first application of the scattering approach to non-trivial geometries 
and non-ideal reflectors can be found in \cite{MaiaNeto05,Rodrigues06,Rodrigues06B}. 
An historial overview of the various forms of the multiple scattering method
can be found in \cite{Milton08}. 
In this paper, we analyse in a detailed manner the full interplay
between finite conductivity, finite temperature and plane-sphere geometry, 
which was only briefly discussed in \cite{Canaguier10}. 

This paper is organized as follows: our main goals and  basic notations are presented in Sec. II. 
In Sec. III, we develop the scattering approach for the plane-sphere geometry and derive the formal 
results used in Secs. IV (perfect reflectors) and V (plasma and Drude models). 
Concluding remarks are presented in Sec. VI.

\section{Definitions and general outlook}

We consider the Casimir interaction between a metallic sphere of radius $R$ and a metallic  plate at a distance of closest approach $L$
at an arbitrary temperature $T,$
as shown in Fig.~1. The center-to-plate distance is
\[
{\cal L} = L+R.
\]
Experimental results for the Casimir force in the plane-sphere geometry are usually compared with PFA-based theoretical models \cite{Derjaguin68}.
The spherical surface is assumed to be nearly flat over a scale of the order of $L,$ and then the total Casimir energy is obtained by adding
the contributions corresponding to different local inter-plate distances over the sphere surface. Although the Casimir energy is not additive,
PFA is usually expected to provide an accurate description when $R\gg L.$
The resulting Casimir force $F^{\rm PFA}$ is
\begin{equation}
\label{PFA}
F^{\rm PFA} = 2\pi R\, \frac{{\cal F}_{\rm PP}}{A}.
\end{equation}
${\cal F}_{\rm PP}$ is the Casimir free energy for two parallel plates of area $A.$
Thus, PFA  neglects diffraction due to the sphere curvature and provides  a  direct connection to the  parallel-plates geometry, the force being proportional to the Casimir free energy per unit area calculated for this
much simpler geometry.

\begin{figure}[h]
\centering
%\vspace{4cm}
\includegraphics[width=3cm]{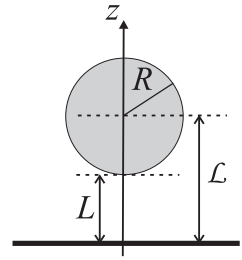}
\caption{Sphere of radius $R$ and a flat plate at
a distance $L$, with center-to-plate distance $\cL\equiv L+R.$}
\end{figure}

In this paper, we develop an exact theoretical model taking diffraction fully into account. We first compute the
Casimir free energy ${\cal F}(L,T)$ for the plane-sphere geometry from the scattering formula \cite{LambrechtNJP06}
(Sec.~III) and then derive the force $F$ and entropy $S$ from
\begin{equation}
\label{FG}
F = -\frac{\partial {\cal F}}{\partial L}, \;\;\;\;\; S = - \frac{\partial {\cal F}}{\partial T}.
\end{equation}
%The latter is the relevant  quantity in frequency-shift measurements, as in the experiment performed by Krause {\it et al} \cite{Krause07} to test the validity of PFA.

The latter is at the core of the ongoing debate about a possible violation of the third law of thermodynamics  in  the  dissipative Drude model \cite{Klim06, Milton08R}. Negative Casimir entropy values found for parallel plates have been explained in terms of the coupling to a heat bath associated to dissipation  \cite{Ingold09}. But here we find negative entropies even for the perfect reflector model, showing that  geometry itself  plays a non-trivial thermodynamical role, provided that beyond-PFA diffraction effects are taken into account.

To quantify the deviation from PFA, we define the quantity
\begin{equation}
\label{rho}
\rho_F = \frac{F}{F^{\rm PFA}}.
\end{equation}

The ratio
\begin{equation}
\label{thetadef}
\vartheta =  \frac{F(L,T)}{F(L,0)}
\end{equation}
represents the temperature correction at a given separation distance and for a given model.
In the next sections, we calculate $\rho_F$ and $\vartheta$ for the perfect reflector, plasma and Drude models for metallic surfaces.

The Drude dielectric function
\begin{equation}
\label{Drude}
\epsilon(i\xi)= 1+ \frac{\omega_P^2}{\xi(\xi+\gamma)}
\end{equation}
at imaginary frequencies $\omega=i\xi$
contains two frequency scales: the plasma frequency $\omega_P$ and the relaxation frequency $\gamma.$  The  plasma
dielectric function  is obtained from (\ref{Drude}) in the lossless limit $\gamma\rightarrow 0.$
Note, however, that there is no continuity in the Casimir force from the Drude to the plasma model in this limit \cite{Ingold09}.
By taking the further limit $\omega_P\rightarrow \infty$ from the plasma model, we recover the  perfect reflector  limit (to be discussed in Sec. IV),
 which corresponds to an
infinite dielectric function at all frequencies.

In Sec. V, we provide a detailed comparison between plasma and Drude models for the Casimir effect.
Since the dc conductivity $\sigma_0=\omega_P^2/\gamma$ diverges in the limit $\gamma\rightarrow 0,$
the Drude model is expected, in principle, to provide a more realistic description of normal (i.e., non-superconducting) metals.
However, experimental data are surprisingly in better agreement with the plasma model when PFA is employed to analyze the plane-sphere
geometry \cite{DeccaPRD07}.
Here we show that the results from the two models are actually closer than predicted by PFA-based theories, which might help to solve this paradox.
We consider a rich parameter space containing five different length scales: besides the geometrical scales $L$ and $R$, our problem contains the
thermal wavelength ($k_B$ is the Boltzmann constant)
\[
\lambda_T= \frac{\hbar c}{k_B T}
\]
the plasma wavelength $\lambda_P=2\pi c/\omega_P$ and the wavelength corresponding to the relaxation frequency $\lambda_{\gamma}=2\pi c/\gamma.$
Different orderings of these length scales can in principle be considered, leading to various regimes associated to non-trivial interplays between temperature,
geometry, finite conductivity and dissipation. Most experiments are performed with gold at room temperature, hence the numerical results presented
in the following sections
 correspond to
 $\lambda_T=7.6\,\mu{\rm m},$
$\lambda_P=136\,{\rm nm}$ (plasma and Drude) and $\lambda_{\gamma}/\lambda_P=250$ (Drude).

\section{Scattering approach in the plane-sphere geometry}

In this section,  we apply the scattering approach to the plane-sphere geometry (see Fig.~1)
 at finite temperature. The Casimir free energy is written as a sum over the Matsubara
 frequencies $\xi_n$ ($n\ge 0$) \cite{LambrechtNJP06}:
\begin{eqnarray}
\label{depart}
&&\cF=k_B T \sum_n^{'}\log{\rm det}
\left(1- \cM (\xi_n)\right) \;,\; \xi_n=\frac{2\pi n k_B T}\hbar  \nonumber \\
&&\cM(\xi)\equiv \cR_\S(\xi) e^{-\cK(\xi) \cL} \cR_\P(\xi)
e^{-\cK(\xi) \cL},
\end{eqnarray}
where the primed sum means that the $n=0$ term is counted for a half. The reflection operators of the sphere,  $\cR_\S(\xi)$, and the plate, $\cR_\P(\xi)$, are evaluated with reference points at the sphere center and at its projection on the plane, respectively. The operator  $e^{-\cK(\xi) \cL}$ accounts for  one-way propagation
along the $z$ axis between these  points, separated by the length $\cL.$
Thus, the operator $\cM (\xi)$  represents one round-trip propagation inside the open cavity formed by the two surfaces.

The plane-wave basis  $|\bk,\pm,p \rangle_{\xi}$ ($\bk =$
wavevector component parallel to the $xy$ plane, $p=\TE,\TM$ for polarization and
$+/-$ for upwards/downwards propagation direction) is well adapted to the description
of the propagation  operator $e^{-\cK(\xi) \cL},$ which is diagonal in this basis with matrix elements
$e^{-\kappa \cL},$  $\kappa = \sqrt{\xi^2/c^2+k^2}$ representing the wave-vector $z$-component associated to the imaginary frequency $\xi$.
Reflection on the plane also preserves all plane wave quantum numbers but the propagation direction, and
the non zero elements of  ${\cal R}_\P(\xi)$
are given by the standard  Fresnel specular reflection amplitudes $r_p(\bk,\xi)$ for an homogenous medium.

On the other hand, the multipole basis $|\ell m P\rangle_{\xi}$, with
$\ell(\ell+1)$ and $m$ denoting the usual angular momentum eigenvalues
(with $\ell=1,2,...$, $m=-\ell,...,\ell$)
and $P=E,M$ representing electric and magnetic multipoles,
is well adapted to the spherical symmetry of ${\cal R}_\S(\xi)$.
By rotational symmetry around the $z$-axis, $\cM(\xi)$
commutes with the angular momentum operator $J_z$.
Hence $\cM(\xi)$ is block diagonal, and each block $\cM^{(m)}(\xi)$ (corresponding to a given subspace $m$)
yields an independent contribution to the Casimir energy.
We find its matrix elements in the multipole basis after introducing the  spectral resolution of the identity operator
in the plane wave basis:
\begin{eqnarray}
\label{Mintegral}
\nonumber
\cM^{(m)}(\xi)_{1,2} & = &
\int\frac{d^2\bk}{(2\pi)^2}\sum_{p=\TE,\TM}
\langle \ell_1 m P_1 | {\cal R}_S(\xi) |\bk,+,p \rangle\\
&& \times  r_p(\bk,\xi)e^{-2\kappa {\cal L}} \,\langle \bk,-,p|\ell_2 m P_2\rangle .
\end{eqnarray}
This expression has a simple interpretation when read from right to left:
a multipole wave $|\ell_2 m  P_2\rangle$ is first decomposed into plane waves
(coefficients $\langle \bk,-,p|\ell_2 m P_2\rangle$) which
propagate towards the plane (factor $e^{-\kappa {\cal L}}$). After reflection by the plane
(specular amplitude  $r_p(\bk,\xi)$), the plane wave components
 propagate back to the sphere (second factor  $e^{-\kappa {\cal L}}$) and are finally scattered
into a new multipole wave $|\ell_1 m P_1\rangle.$

The matrix elements of ${\cal R}_\S$
 in (\ref{Mintegral}) represent the multipole components of the  field scattered by the sphere
for a given incident plane wave. In Mie scattering calculations, one usually assumes that
the incident plane wave  propagates along the $z$-direction \cite{Bohren}. Here this is no longer possible, since
we have to consider  field modes propagating simultaneously along all possible directions. It is then useful to re-formulate \cite{Mazolli03} the Mie
scattering expressions  in terms of the matrix elements of  finite rotation \cite{Varshalovich}
\[
d^{\ell}_{m,m'}(\theta) = \langle \ell m | e^{-i\theta J_y} |\ell m'\rangle
\]
with $m'=\pm 1$ accounting for the photon spin.

The resulting expressions for $\langle \ell_1 m P_1 | {\cal R}_S(\xi) |\bk,+,p \rangle$ (see appendix), are proportional to the Mie coefficients $a_{\ell}(i\xi)$  and  $b_{\ell}(i\xi)$  \cite{Bohren}, which represent the scattering amplitudes for electric and magnetic multipole waves. At the imaginary frequency axis, they are written in terms of the modified Bessel functions \cite{Abramowicz} evaluated at the `size parameter'
\[
{\tilde \xi}= \frac{\xi R}{c}
\]
 as follows:
\begin{eqnarray}
\label{Mie_geral1}
a_{\ell}(i \tilde{\xi}) & = &  \frac{\pi}{2} \,  \frac{n^2 s_{\ell}^{(a)} - s_{\ell}^{(b)}}{n^2 s_{\ell}^{(c)} - s_{\ell}^{(d)}} \\
\label{Mie_geral2}
b_{\ell}(i \tilde{\xi}) & = &  \frac{\pi}{2} \, \frac{ s_{\ell}^{(a)} - s_{\ell}^{(b)}}{s_{\ell}^{(c)} - s_{\ell}^{(d)}} \\
\nonumber
s_{\ell}^{(a)} & = &  I_{\ell+1/2}(n \tilde{\xi}) \left(  I_{\ell+1/2}(\tilde{\xi}) - \tilde{\xi} I_{\ell-1/2}(\tilde{\xi}) \right) \\
\nonumber
s_{\ell}^{(b)} & = &  I_{\ell+1/2}(\tilde{\xi}) \left(  I_{\ell+1/2}(n \tilde{\xi}) - n \tilde{\xi} I_{\ell-1/2}(n \tilde{\xi}) \right) \\
\nonumber
s_{\ell}^{(c)} & = &  I_{\ell+1/2}(n \tilde{\xi}) \left(  K_{\ell+1/2}(\tilde{\xi}) + \tilde{\xi} K_{\ell-1/2}(\tilde{\xi}) \right) \\
\nonumber
s_{\ell}^{(d)} & = &  K_{\ell+1/2}(\tilde{\xi}) \left(  I_{\ell+1/2}(n \tilde{\xi}) - n \tilde{\xi} I_{\ell-1/2}(n \tilde{\xi}) \right)
\end{eqnarray}
with $n=\sqrt{\epsilon}$ representing the sphere refractive index.

 We also derive in the appendix
 the change-of-basis matrix elements $ \langle \bk,-,p|\ell_2 m P_2\rangle.$
They yield, when replaced into (\ref{Mintegral}),
explicit expressions for the matrix elements of the round-trip operator  $\cM^{(m)}(\xi),$
which we organize as a block matrix:
\begin{equation}
\label{blocks}
\cM^{(m)}(\xi) = \left(
\begin{array}
[c]{cc}%
\cM^{(m)}(E,E)
 & \cM^{(m)}(E,M)
\\
\cM^{(m)}(M,E) & \cM^{(m)}(M,M)
\end{array}
\right).
\end{equation}
Each block is the sum of TE and TM contributions:
\(
\cM^{(m)}(P_1,P_2)=\sum_p \cM^{(m)}_{p}(P_1,P_2).
\)
For the diagonal blocks, we find
\begin{eqnarray}\label{diagonal}
& \cM^{(m)}_{\TE}(E,E)_{\ell_1,\ell_2} =
 \sqrt{\frac{(2\ell_1+1)\pi}{\ell_2(\ell_2+1)}}\,A^{(m)}_{\ell_1,\ell_2,\TE}\,a_{\ell_1}(i\xi) \\
\label{diagonalTM}
& \cM^{(m)}_{\TM}(E,E)_{\ell_1,\ell_2} =
 \sqrt{\frac{(2\ell_1+1)\pi}{\ell_2(\ell_2+1)}}\,B^{(m)}_{\ell_1,\ell_2,\TM}\,a_{\ell_1}(i\xi) \\
&\cM^{(m)}_{\TM}(M,M)_{\ell_1,\ell_2} =
 \sqrt{\frac{(2\ell_1+1)\pi}{\ell_2(\ell_2+1)}}\,A^{(m)}_{\ell_1,\ell_2,\TM}\,b_{\ell_1}(i\xi) \\
\label{diagonalMM}
& \cM^{(m)}_{\TE}(M,M)_{\ell_1,\ell_2} =
 \sqrt{\frac{(2\ell_1+1)\pi}{\ell_2(\ell_2+1)}}\,B^{(m)}_{\ell_1,\ell_2,\TE}\,b_{\ell_1}(i\xi) .
\end{eqnarray}

The matrices $A^{(m)}$ and $B^{(m)}$ do neither depend on the radius $R$ nor on the refractive index of the sphere
(spherical harmonics $Y_{\ell m}(\theta,\varphi)$   \cite{Varshalovich} calculated at the azimuthal angle $\varphi=0$):
\begin{eqnarray}\label{A}
 A^{(m)}_{\ell_1,\ell_2,p} = -i m
\int_0^{\infty}\frac{dk}{\kappa}[d^{\ell_1}_{m,1}(\theta^{(+)})+
d^{\ell_1}_{m,-1}(\theta^{(+)})]
\\
\nonumber
\times Y_{\ell_2m}(\theta^{-})\, r_p(k)\, \exp(-2\kappa {\cal L})
\\
\label{B}
 B^{(m)}_{\ell_1,\ell_2,p} =  -\frac{c}{\xi}
\int_0^{\infty} dk \frac{k}{\kappa}[d^{\ell_1}_{m,1}(\theta^{(+)})-
d^{\ell_1}_{m,-1}(\theta^{(+)})]
\\
\nonumber
\times \partial_{\theta} Y_{\ell_2m}(\theta^{-})\, r_p(k)\, \exp(-2\kappa {\cal L}) \\
\label{theta_complex}
\sin\theta^{\pm}=-i\frac{ck}{\xi},\;\;\cos\theta^{\pm}=\pm\frac{c\kappa}{\xi},\;\;\kappa\equiv\sqrt{\xi^2/c^2+k^2} .
\end{eqnarray}

Similar expressions are found for the non-diagonal blocks:
\begin{eqnarray}
\nonumber
& \cM^{(m)}_{\TE}(E,M)_{\ell_1,\ell_2} =
i \sqrt{\frac{(2\ell_1+1)\pi}{\ell_2(\ell_2+1)}}\,C^{(m)}_{\ell_1,\ell_2,\TE}\,a_{\ell_1}(i\xi) \\
\nonumber
& \cM^{(m)}_{\TM}(E,M)_{\ell_1,\ell_2} =
i \sqrt{\frac{(2\ell_1+1)\pi}{\ell_2(\ell_2+1)}}\,D^{(m)}_{\ell_1,\ell_2,\TM}\,a_{\ell_1}(i\xi) \\
\nonumber
&\cM^{(m)}_{\TM}(M,E)_{\ell_1,\ell_2} =
- i \sqrt{\frac{(2\ell_1+1)\pi}{\ell_2(\ell_2+1)}}\,C^{(m)}_{\ell_1,\ell_2,\TM}\,b_{\ell_1}(i\xi) \\
\nonumber
& \cM^{(m)}_{\TE}(M,E)_{\ell_1,\ell_2} =
- i \sqrt{\frac{(2\ell_1+1)\pi}{\ell_2(\ell_2+1)}}\,D^{(m)}_{\ell_1,\ell_2,\TE}\,b_{\ell_1}(i\xi) .
\end{eqnarray}
 $C^{(m)}$ and $D^{(m)}$ are also written in terms of spherical harmonics and rotation matrices:
\begin{eqnarray}\label{C}
 C^{(m)}_{\ell_1,\ell_2,p} =   \frac{c}{\xi}
\int_0^{\infty}dk \frac{k}{\kappa}[d^{\ell_1}_{m,1}(\theta^{(+)})+
d^{\ell_1}_{m,-1}(\theta^{(+)})]
\\
\nonumber
\times \partial_{\theta} Y_{\ell_2m}(\theta^{-})\, r_p(k)\, \exp(-2\kappa {\cal L})
\\
\label{D}
 D^{(m)}_{\ell_1,\ell_2,p} =  i  m
\int_0^{\infty} \frac{dk}{\kappa}[d^{\ell_1}_{m,1}(\theta^{(+)})-
d^{\ell_1}_{m,-1}(\theta^{(+)})]
\\
\nonumber
\times  Y_{\ell_2m}(\theta^{-})\, r_p(k)\, \exp(-2\kappa {\cal L}) .
\end{eqnarray}

By inspection of (\ref{A}) and (\ref{B}), it is easy to show that the diagonal blocks are invariant when we replace $m\rightarrow -m,$ whereas
the non-diagonal blocks given by (\ref{C}) and (\ref{D}) change sign. Thus, the matrix $\cM^{(0)}$ is block diagonal and
 $\det(1-\cM^{(m)}(\xi_n))$ does not
depend on the sign of $m,$ allowing us to write the Casimir free energy as a double primed sum (with only non-negative values of $m$):
\begin{equation}
\label{double}
\cF=2 k_B T \sum_n^{'}\sum_m^{'}\log{\rm det}
\left(1- \cM^{(m)} (\xi_n)\right) .
\end{equation}
 The zero-temperature limit can also
be obtained from (\ref{double}) by changing the sum over Matsubara frequencies by an integral over  $\xi:$
\begin{equation}
\label{zerotemp}
k_B T \sum_n^{'} \rightarrow \hbar \int_0^{\infty} \frac{d\xi}{2\pi}\quad\quad\quad\quad (T\rightarrow 0) .
\end{equation}

{\it Numerical  considerations.}
When evaluating the free energy from (\ref{double}),
one  needs to truncate the vector space at some maximum value $\ell_{\rm max}$ of angular momentum.
From the localization principle \cite{Nussenzveig}, the value of $\ell_{\rm max}$ required for a given accuracy level
is expected to scale with the size parameter $\tilde \xi$
which captures the dependence of $\cM^{(m)}$ on the sphere radius according to (\ref{Mie_geral1})-(\ref{Mie_geral2}).
Characteristic frequencies, giving the main contribution to the Casimir energy, scale as $\xi \sim c/L$.
As a consequence, the required $\ell_{\rm max}$
 should scale as $R/L$ for intermediate and short separation distances. This is verified by the numerical evaluations discussed below. They are limited
to $\ell_{\rm max}=85$ at the moment ($\ell_{\rm max}=45$ for Drude model), allowing us to calculate for $R/L <20$ ($R/L <10$ for Drude model), not far from for the experimental range \cite{Krause07}
$R/L\gtrsim10^2$  which would require a larger $\ell_{\rm max}.$

\begin{figure}[t]
\centering
\includegraphics[width=8cm]{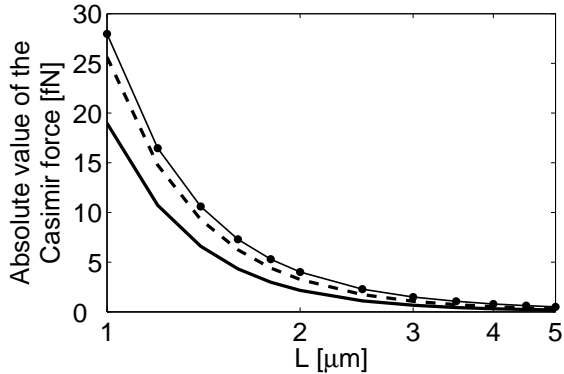}
\caption{ Casimir force as a function of distance $L$  at $T=300$K
for a sphere radius $R=10\,\mu{\rm m},$
computed for perfect reflectors (dashed) or gold surfaces described
by the Drude model (solid). The dotted line represents the PFA result for perfect reflectors.}
\label{fig_energy}
\end{figure}

Apart from this restriction, we are able to calculate the exact Casimir free energy and force
for parameters closely mimicking the experimental conditions. As a first example, we plot in Fig.~\ref{fig_energy} the Casimir force as a function of
$L$ for a sphere of radius $R=10\,\mu{\rm m}$ at room temperature. The  dashed and solid lines correspond to the perfect reflector model
(to be discussed in detail in Sec.~IV) and to the Drude model (Sec.~V), respectively.
We also show the PFA result for perfect reflectors (dotted line). The  interplay between the effects of temperature, geometry, finite conductivity and dissipation is better
 understood when considering the ratios defined in Sec.~II, which are analyzed in detail in the next sections. At the limit of large separation distances,
 simple analytical results can be derived, as discussed below.

{\it Large-distance limit.} When $R \ll \cL,$ the characteristic size parameters scale as
${\tilde \xi}\sim R/{\cal L} \ll 1$ and the Mie coefficients  $a_{\ell}(i\xi)\sim {\tilde \xi}^{2\ell+1}$, with the magnetic coefficients $b_{\ell}(i\xi)$
of the same order or smaller than $a_{\ell}(i\xi)$
   depending on the material properties of the sphere. The resulting matrix elements $\cM^{(m)}(P,P')_{\ell_1,\ell_2}$ are very
small and the dominant contribution comes from $\ell=1$ ($m=0,1$), which corresponds to the dipole contributions:
\begin{equation}
\label{det_ld}
\log\det(1-\cM^{(m)})\approx - \sum_{P=E,M}\cM^{(m)}(P,P)_{1,1}.
\end{equation}
The explicit expressions for the electric dipole matrix elements are obtained from (\ref{diagonal})-(\ref{diagonalTM}) and
(\ref{A})-(\ref{B}):
\begin{eqnarray}
\label{dipeletr0}
\cM^{(0)}(E,E)_{1,1}& =    - \frac{3a_1(i\xi)}{2(\xi/c)^3}\int_0^{\infty} dk & \frac{k^3}{\kappa}r_{\rm TM}e^{-2\kappa {\cal L}}   \\
\label{dipeletr1}
\cM^{(1)}(E,E)_{1,1}& =     \frac{3a_1(i\xi)}{4(\xi/c)^3}\int_0^{\infty} dk
&\Bigl(
\frac{k\xi^2}{c^2\kappa}r_{\rm TE}\\
\nonumber
 && -k\kappa\, r_{\rm TM}\Bigr) e^{-2\kappa \cL} .
\end{eqnarray}
The magnetic dipole elements are obtained from (\ref{dipeletr0})-(\ref{dipeletr1}) by replacing $a_1(i\xi)\rightarrow b_1(i\xi)$ and interchanging
$r_{\rm TM}\leftrightarrow r_{\rm TE}.$ When the refractive index is finite,
its contribution is negligible and then the dominant contribution is the electric dipole one (Rayleigh scattering).
To proceed further, we need to specify the material properties for the sphere and plane and the corresponding
Mie  and Fresnel coefficients appearing in (\ref{dipeletr0})-(\ref{dipeletr1}). In the next two sections, we consider  perfect reflectors, plasma and Drude metals.

\section{Perfect reflectors}

Within the perfect reflector model for metals, the dielectric permittivity is taken to be infinite at all frequencies. This  simple albeit unphysical
model for metals  provides an accurate description at large separations  in the ideal zero-temperature case.
At finite temperatures, however, it is still unclear if the perfect reflector model reproduces the correct long distance regime for real metals. In fact,
for parallel plates, it predicts a
force twice as large as the value obtained within the dissipative Drude model for metals in the long-distance limit,
while we would in principle expect  the two results to agree in this limit.

 It is thus extremely important to compare the results
obtained from the different models for the plane-sphere geometry. In this section, we start with the  perfect reflector model, with
 Fresnel reflection coefficients $r_{\rm TE}^\perf=-r_{\rm TM}^\perf=-1.$ The Mie coefficients are
 obtained by taking $n\gg 1$ and $n{\tilde \xi}\gg 1$ in (\ref{Mie_geral1})-(\ref{Mie_geral2}):
\begin{eqnarray}
\label{Mie_a}
&& a_{\ell}^\perf(i\xi)= \frac{\pi}{2}(-)^{\ell+1}
\frac{  \ell I_{\ell+1/2}(\tilde \xi)-\tilde \xi  I_{\ell-1/2}(\tilde \xi)}
{ \ell K_{\ell+1/2}(\tilde \xi)+\tilde \xi  K_{\ell-1/2}(\tilde \xi)}
\\
\label{Mie_b}
&& b_{\ell}^\perf(i\xi)= \frac{\pi}{2}(-)^{\ell+1}
\frac{  I_{\ell+1/2}(\tilde \xi)}
{  K_{\ell+1/2}(\tilde \xi)}.
\end{eqnarray}
The Fresnel and Mie coefficients written above  can also be obtained from the plasma model expressions (discussed in the next section) by taking the
limit $\lambda_P\rightarrow 0.$

{\it Large-distance limit for perfect reflectors.} For ${\tilde \xi}\ll 1$ we may take the power expansion of (\ref{Mie_a})-(\ref{Mie_b}):
\begin{equation}
\label{Mie_expansion}
a_1^\perf(i{\tilde \xi}) = -\frac{2}{3}\,{\tilde \xi}^3 + {\cal O}({\tilde \xi})^5,\;\;\; b_1^\perf(i{\tilde \xi})= \frac{1}{3}\,{\tilde \xi}^3 + {\cal O}({\tilde \xi})^5.
\end{equation}
Note that $b_1,$
representing the magnetic dipole contribution, is of the same order of the electric dipole coefficient $a_1$ for ${\tilde \xi}\ll 1.$  This property holds
whenever $n {\tilde \xi}\gg 1,$ which is also the case for the plasma model at low frequencies when $\lambda_P\ll R.$
On the other hand, for any finite dielectric constant, the magnetic dipole is always of the order of the electric quadrupole and thus
 much smaller than the electric dipole contribution in the low-frequency limit. We find this so-called Rayleigh scattering regime when discussing the long-distance limit within the
 Drude model in the next section.

We insert (\ref{Mie_expansion}) and the values for the Fresnel coefficients into (\ref{double})-(\ref{dipeletr1}) and write the resulting
expression for the free energy in terms of the thermal wavelength~$\lambda_T:$
\begin{eqnarray}
\label{large-distance_pr1}
\mathcal{F}^\perf &  = & - \frac{3 R^3}{4 \mathcal{L}^3} \frac{\hbar c}{\lambda_T} \sum_n^{'} \left( 1 +  2 \nu n + 2 \nu^2 n^2 \right) e^{-2 \nu n}  \\
\nu&  = & \frac{2 \pi \mathcal{L}}{\lambda_T}.
\end{eqnarray}
The evaluation of the sum over Matsubara frequencies in (\ref{large-distance_pr1}) is straightforward:
\begin{eqnarray}
\label{Eperfectanalytical}
&&\cF^\perf =
-\frac{3 \hbar c R^3}{4 \lambda_{T} \cL^3}\, \phi(\nu), \quad\quad\quad \cL \gg R\\
&&\phi(\nu) \equiv \frac{ \nu\sinh\nu + \cosh\nu ( \nu^2 + \sinh^2\nu )}
{2 \sinh^3\nu}. \nonumber
\end{eqnarray}
The low temperature  approximation is  derived  by expanding $\phi(\nu)$  in powers of $\nu:$
\begin{equation}
\label{low-temperature}
\cF^\perf  \approx
- \frac{9 \hbar c R^3}{16 \pi \cL^4} \left( 1 - \frac{\nu^4}{135} + \frac{4\nu^6}{945} \right), \lambda_T \gg  \cL \gg R.
\end{equation}
Note that the zero-temperature limit contained in the above expression
can also be obtained by replacing the sum over $n$ in (\ref{large-distance_pr1}) by an integral over $\xi$ as in (\ref{zerotemp}).

The first finite-temperature correction in  (\ref{low-temperature}) does not depend on $\cL,$ so that the temperature correction to the Casimir force
comes from the next-to-leading-order term, proportional to $(k_BT)^6.$ It is a repulsive contribution from thermal photons that makes the net force slightly less
attractive. We discuss this point further in the context of the numerical evaluations presented below.

We can  compute the high-temperature limit by taking $\nu\rightarrow\infty$ in (\ref{Eperfectanalytical}) or  by considering the
$n=0$ Matsubara frequency contribution in (\ref{large-distance_pr1}):
\begin{equation}
\label{high-T_pr}
\cF^\perf \approx -\frac{3 \hbar c R^3}{8 \lambda_T \cL^3}
\quad,\quad \cL \gg \lambda_T ,\, R .
\end{equation}

It is also interesting to compute the Casimir entropy from (\ref{Eperfectanalytical}):
\begin{eqnarray}
\label{Sperfectanalytical}
&&S^\perf = \frac{3 k_\mathrm{B} R^3}{4 \cL^3}
\left( \phi(\nu) + \nu \phi^\prime(\nu) \right),\quad \cL \gg R.
\end{eqnarray}
Remarkably, this expression yields $S^\perf<0$  for  $\nu\lesssim 1.5$, that is $L\lesssim 1.8\mu$m at $T=300$K. In the parallel plates geometry,  negative entropies have been found for the dissipative Drude model only \cite{Bezerra02}. Here we find negative entropies also for perfect reflectors over a wide separation distance range,
as discussed below.

\begin{figure}[t]
\centering
\includegraphics[width=8cm]{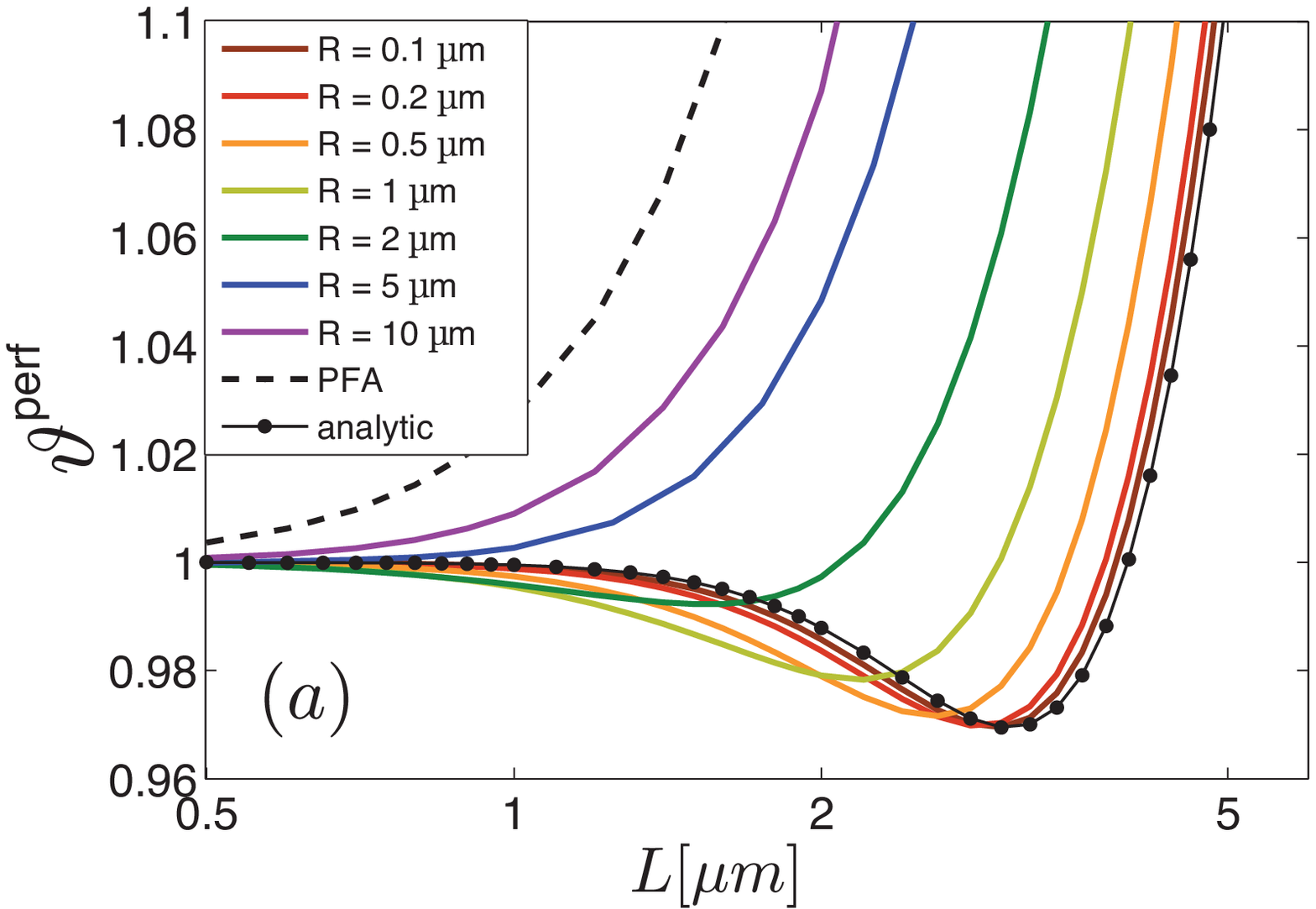}
\includegraphics[width=8cm]{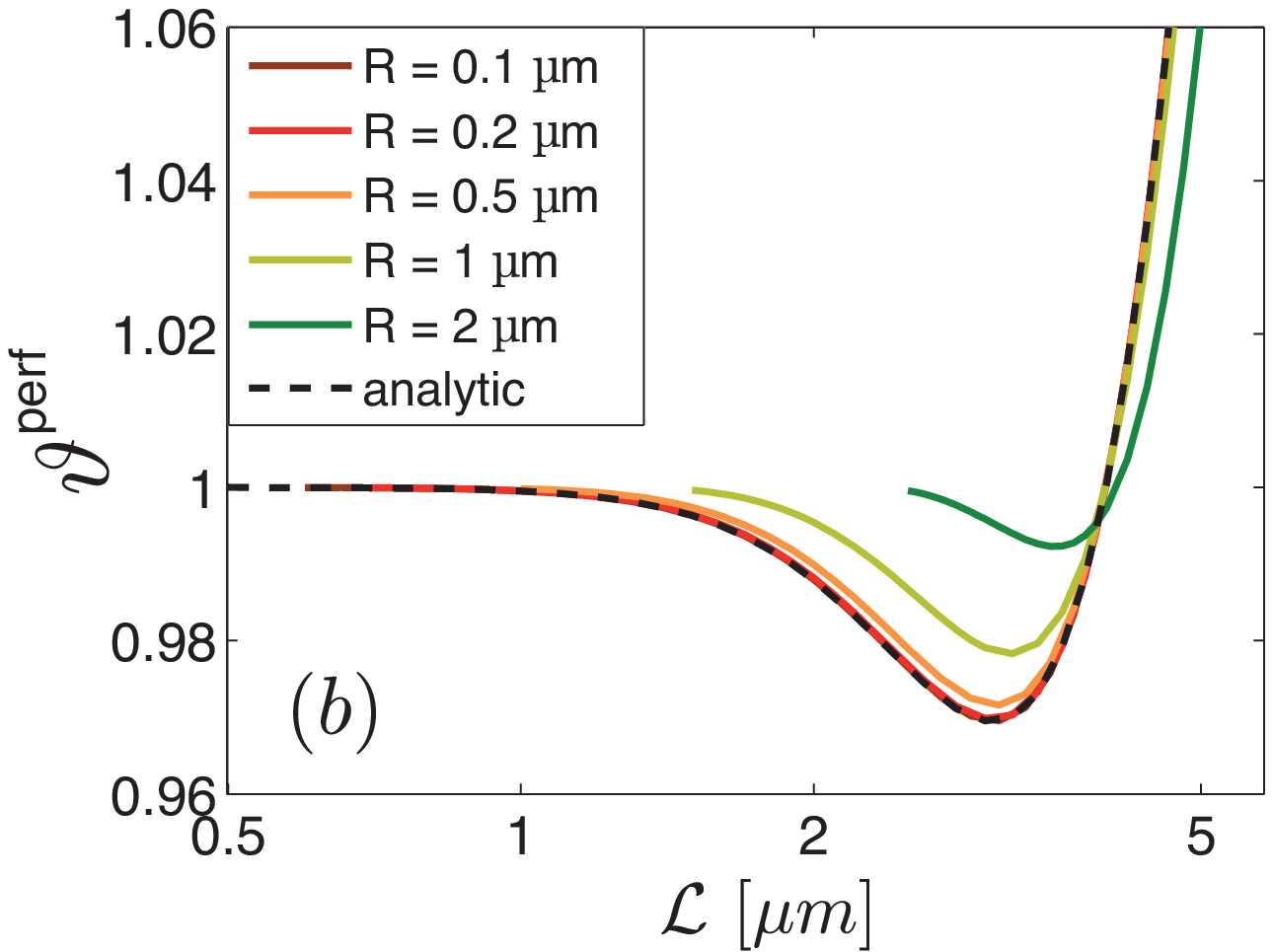}
\caption{Thermal Casimir force at $T=300$K computed between perfectly reflecting
sphere and plane, divided by the zero temperature force, as function of surface distance $L$ (a) and distance-to-center $\cL$ (b).
 Solid lines from bottom
to top correspond to increasing values of sphere radii.
The upper dashed curve represents the PFA result while the lower dotted
curve corresponds to the analytical large-distance expression.
[Colors online].} \label{theta_perf}
\end{figure}

{\it Numerical results.}
In Fig.~\ref{theta_perf}a, we plot the  ratio $\vartheta^\perf$ [see (\ref{thetadef})], quantifying the thermal correction to the Casimir force, for different sphere radii  as a function of the separation distance $L.$
We also show the results obtained by using the PFA formula  (\ref{PFA})  (dashed line) and by deriving the analytical long-distance expression (\ref{Eperfectanalytical})
with respect to $L$ after replacing $\cL\approx L$ (dotted line).
The agreement with  the analytical formula is as expected better if we keep the variable $\cL$ in (\ref{Eperfectanalytical}), as shown in Fig.~\ref{theta_perf}b, where
we plot $\vartheta^\perf$ as a function of $\cL.$

At very short  distances $L\ll \lambda_T$ we recover, as expected, the zero temperature result ($\vartheta=1$).
 As the distance
increases, we find, in most cases,  that  $\vartheta$ decreases below one, reaches a radius-dependent minimum and then increases again at long distances.
As long as $R$ is not too large,
the thermal photons provide a repulsive contribution (thus decreasing the magnitude of the overall attractive force) over a distance range that
becomes wider as $R$ decreases.  This range corresponds to
 $L\lesssim\lambda_T/\pi$ when $R\ll L,$  as obtained  from (\ref{Eperfectanalytical}).

\begin{figure}[t]
\centering
\includegraphics[width=8cm]{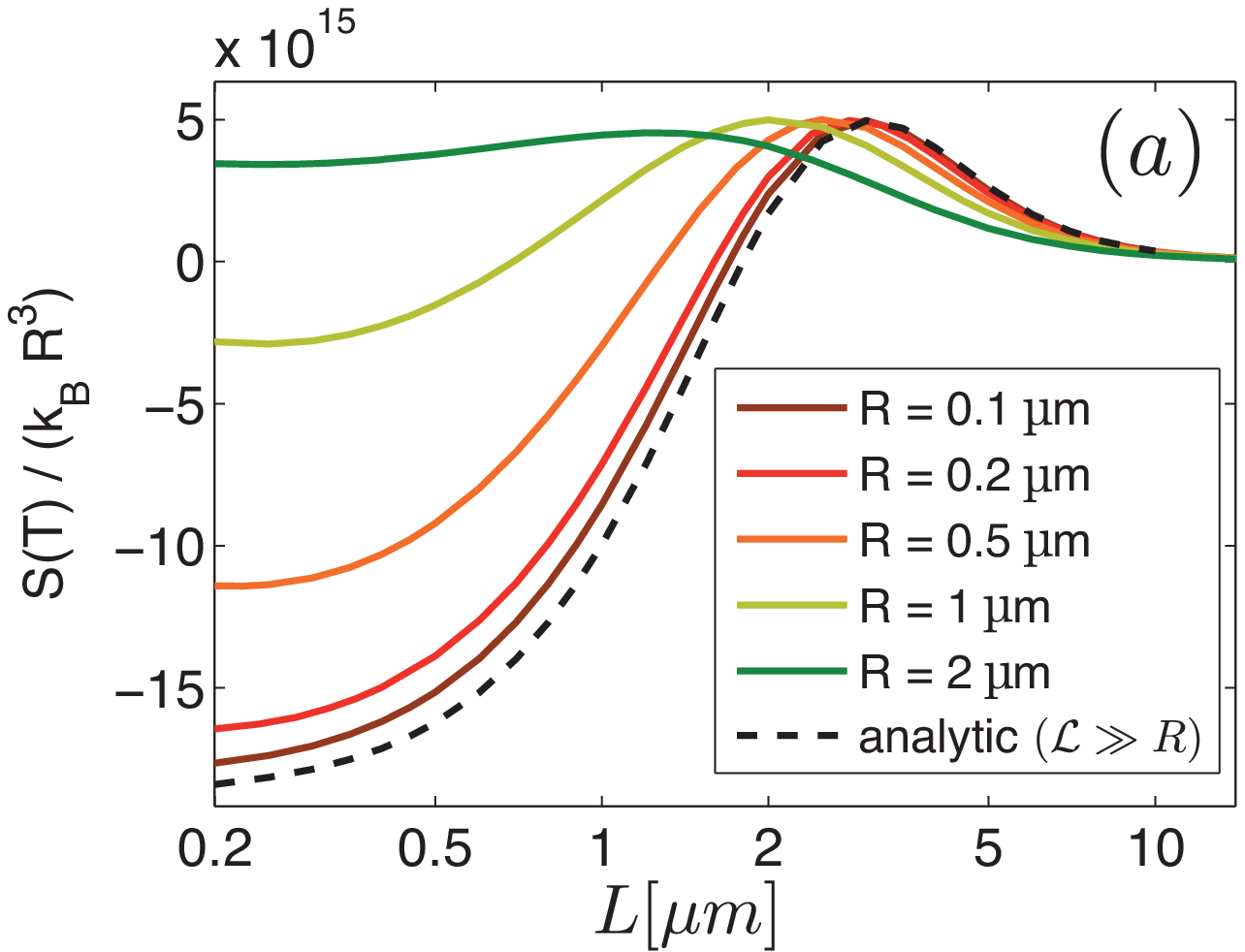}
\includegraphics[width=8cm]{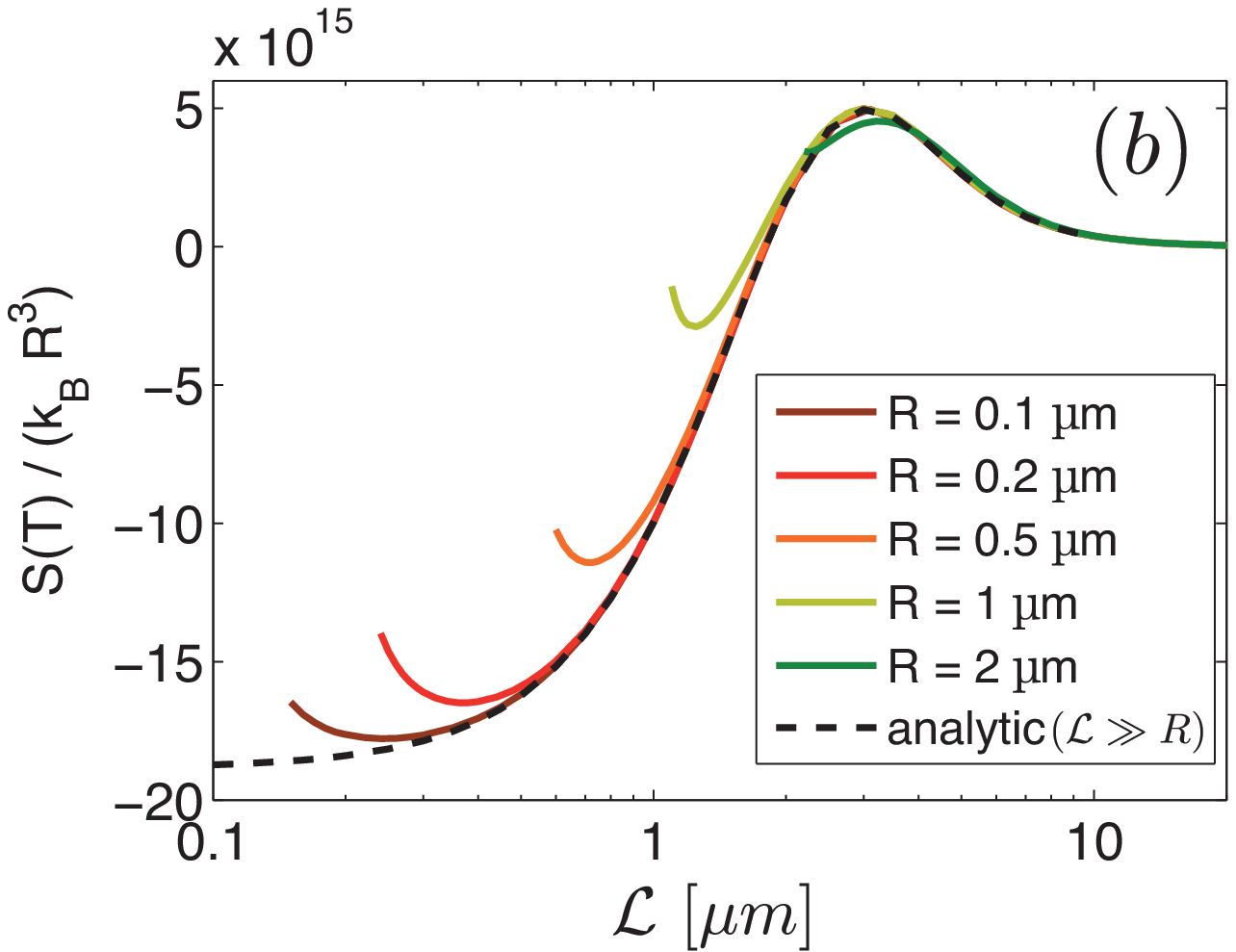}
\caption{Casimir entropy
 at $T=300$K computed for perfectly reflecting
sphere and plane, as a function of surface separation distance $L$ (a) and center-to-plate distance $\cL (b).$
The dashed curve corresponds to the analytical asymptotic expression  for $\cL \gg R.$
[Colors online].} \label{entropy}
\end{figure}

The reduction of the Casimir force is consistent with the negative Casimir entropies found from (\ref{Sperfectanalytical}) in the limit $R\ll L.$
We plot $S$ as a function of $L$ (Fig.~\ref{entropy}a) or $\cL$  (Fig.~\ref{entropy}b) for different sphere radii. Fig.~\ref{entropy}b shows that (\ref{Sperfectanalytical})
provides an accurate description for $\cL/R>4.$ Negative entropies are found for $R$ as large as $1\,\mu{\rm m}.$

An additional relevant property can be inferred from Fig.~\ref{theta_perf}:  the PFA always overestimates the thermal correction for perfect reflectors, the overestimation being
smaller for larger radii (at a given separation distance $L$) as expected. 

\begin{figure}[t]
\centering
\includegraphics[width=8cm]{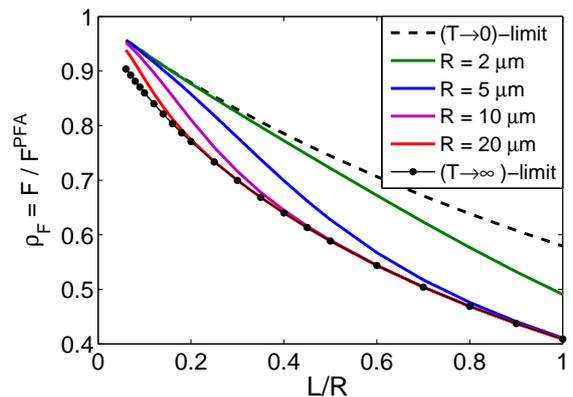}
\caption{
Beyond-PFA correction factor
 computed for perfectly reflecting
sphere and plane at ambient temperature (solid lines) and for asymptotic cases (dashed and dotted lines), as a function of surface separation $L.$
[Colors online].} \label{rho_pr}
\end{figure}

In Fig.~\ref{rho_pr}, we plot the beyond-PFA correction $\rho$ at room temperature as a function of $L/R,$ for different values of $R.$ At zero temperature, the different curves shown in this figure would collapse into a single one
\cite{Emig08,Maia08}.
% close to the solid (black) line corresponding to $R=2\,\mu{\rm m}.$
 For a given ratio $L/R,$ the thermal reduction effect already apparent in Fig.~\ref{theta_perf} is larger for larger radii.
The fact that $\rho_F$ depends on $R$ for a given $L/R$ is again a clear signature of the interplay
between thermal and geometry effects, that can damage the precision of PFA. In the next section, we discuss how this interplay is modified when finite conductivity and dissipation are included in the model.

\section{Plasma and Drude metals}

The plasma model provides the simplest way to take the finite conductivity of metals into account.
The Drude model  is a more accurate description of non-superconducting metals since it also includes the
relaxation of conduction electrons and the associated finite dc conductivity.
In this section, we take the dielectric constant
$\epsilon$ given by (\ref{Drude}) (with $\gamma=0$ in the plasma model) and
 derive analytical and numerical results for the Casimir free energy and force.

{\it Large distance high-temperature limit.}
When $\cL\gg R,$ we take the low-frequency expansion of the Mie coefficients and find, for the plasma model \cite{Tanner84},
\begin{equation}
\label{Mieplasma1}
a_1^\plas \simeq - \frac{2 \txi^3}{3},  \;\;
b_1^\plas \simeq \left( \frac{1}{3}+ \frac{1}{\alpha^2}-
\frac{\coth\alpha}{\alpha} \right) \txi^3
\end{equation}
\begin{equation}
\alpha = \frac{2\pi R}{\lambda_P}.
\end{equation}

Here we also assume that $\cL \gg \lambda_T$ (high-temperature limit), so that we take only the first Matsubara frequency $\xi_0=0$  when computing the Casimir free energy from  (\ref{double})
In the low frequency limit, the Fresnel coefficients are  given by $r_{\rm TE}\approx - r_{\rm TM}\approx -1$, and then we find, from plugging (\ref{Mieplasma1})
into Eqs.~(\ref{det_ld})-(\ref{dipeletr1}) :
\begin{equation}
\label{Mieplasma2}
\cF^\plas \approx  - \frac{3 \hbar c R^3}{8 \lambda_T \cL^3}
\left( 1 + \frac{1}{\alpha^2}-\frac{\coth\alpha}{\alpha} \right) ~ ,~   \cL \gg  \lambda_T, R .
\end{equation}
This result reproduces, as a particular case, the perfectly-reflecting limit given by (\ref{high-T_pr}) when $\lambda_P\ll R.$

For the Drude model,
the TE Fresnel reflection coefficient has the well-known low-frequency limit $r_{\rm TE}\rightarrow 0,$ whereas
the TM coefficient behaves as in the plasma model:  $r_{\rm TM}\approx 1.$
The low-frequency expansion of the Mie coefficients are also quite different from the plasma case.
For the electric dipole coefficient, we find
\begin{equation}
\label{MieDrude}
a_1^\Drud \approx - \frac{2 \txi^3}{3} + \frac{c \txi^4 }{\sigma_0 R}.
\end{equation}
As  discussed in Sec.~IV, the magnetic dipole
 $b_1$ is always much smaller than the electric dipole $a_1$ for any finite value of $\epsilon$ in the limit $\txi\rightarrow 0.$
 For any non-zero value of the relaxation frequency $\gamma$ in (\ref{Drude}), the zero-frequency limit of $\epsilon$ is finite. In contrast with
 the perfect-reflector and plasma cases, the magnetic dipole contribution is then negligible
\begin{equation}
\label{MieDrudeb1}
b_1^\Drud \approx  \frac{\sigma_0 R \txi^4 }{45c}\ll a_1^\Drud
\end{equation}
for any finite value of the  Drude dc conductivity
 $\sigma_0=\omega_P^2/\gamma.$ The Drude sphere at low frequencies thus behaves as an inducible electric dipole
for any finite value of $\sigma_0$, corresponding to the Rayleigh scattering limit.

\begin{figure}[t]
\centering
\includegraphics[width=8cm]{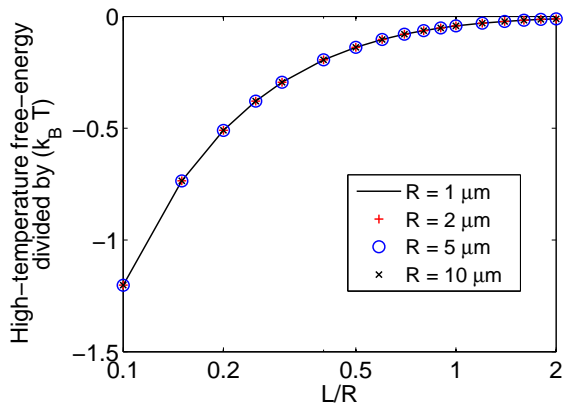}
\caption{
High-temperature Casimir free energy as a function of $L/R$ calculated with the Drude model.
The results do neither depend separately on $L$ and $R$ nor on the material parameters $\lambda_P$ and $\lambda_{\gamma}.$} \label{num_Drude}
\end{figure}

The resulting high-temperature large-distance limit for the free energy reads
\begin{equation}\label{F_Drud}
\cF^\Drud \approx  - \frac{\hbar c R^3}{4 \lambda_T \cL^3} \quad , \quad  \cL \gg  \lambda_T, R .
\end{equation}
Remarkably this result does not depend on the length scales $\lambda_P$ and $\lambda_{\gamma}$ characterizing the material response, whereas
the corresponding plasma result (\ref{Mieplasma2}) clearly depends on $\lambda_P.$
One can show that
this is always the case in the high-temperature limit
$\lambda_T\ll \cL,$ with $\cF^\Drud$ converging to the universal function of $L/R$
shown in Fig.~\ref{num_Drude}, which is
 determined by the contribution of higher multipoles $\ell\le \ell_{\rm max}\sim R/L.$

The expression (\ref{F_Drud})
 corresponds to $\frac{2}{3}$ of the value for perfect reflectors (\ref{high-T_pr}), to be compared with the ratio $\frac{1}{2}$ found in the
parallel-planes geometry \cite{Bostrom00}, which results from the fact that the Fresnel coefficient $r_{\rm TE}$ vanishes at the zero frequency limit.
Here the TE and TM contributions are redistributed into electric and magnetic multipole spherical modes, thus explaining the change from
$\frac{1}{2}$ to $\frac{2}{3}.$ In fact, for perfect reflectors the magnetic dipole contribution proportional to $b_1$ is one-third of the total free energy
(\ref{high-T_pr}), as can be surmised from (\ref{Mie_expansion}) which shows that $|b_1|=|a_1|/2.$
 Since this contribution is negligible  in the Drude model, the free energy is reduced by the factor $\frac{2}{3}.$

\begin{figure}[t]
\centering
\includegraphics[width=8cm]{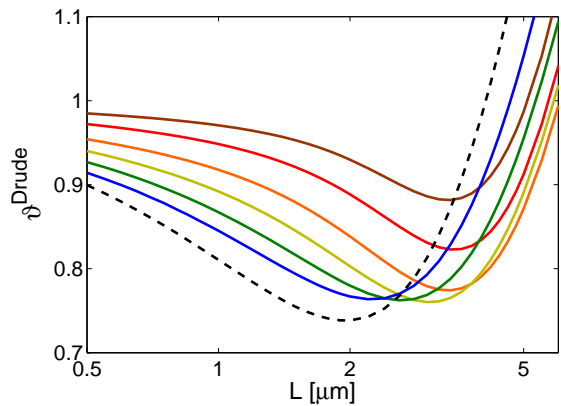}
\caption{Thermal Casimir force correction
computed with the Drude model. The conventions are the
 same  as in Fig.~\ref{theta_perf}.
[Colors online].} \label{Drude1}
\end{figure}

 {\it Numerical results.}
 An important consequence of the discussion presented above
  is  that results from Drude and perfect reflector as well as plasma models  are  closer in the plane-sphere
 geometry than in the parallel-planes geometry.  We have computed the Casimir force numerically for arbitrary (not too small) values of the
 surface  distance $L.$ For the plasma model, the thermal correction $\vartheta^\plas$ (not shown) is found to be close to the values for perfect reflectors
 shown in Fig.~\ref{theta_perf}. On the other hand, the variation of  $\vartheta^\Drud$ calculated within the Drude model
 is remarkably different, as shown in Fig.~\ref{Drude1}.

In contrast with the perfect reflectors and plasma model calculations, for which PFA always overestimates the thermal correction,
PFA underestimates the thermal correction
 at short distances for the Drude model, while it overestimates it at long distances. The overestimation is, however, clearly smaller than for perfect reflectors.
 Since PFA results for plasma metals are  above the values calculated for Drude metals, the exact beyond-PFA Casimir force values at short
 distances for Drude and plasma
 models turn out to be much closer than predicted by the PFA-based theoretical models used in the analysis of experimental data.

\begin{figure}[t]
\centering
\includegraphics[width=8cm]{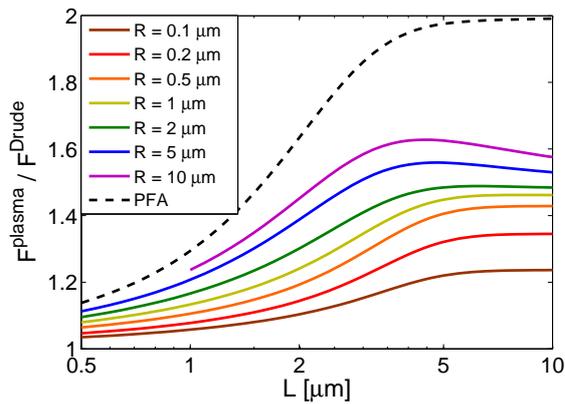}
\caption{Ratio between thermal Casimir force values
calculated with the plasma  and the Drude models,
as a function of surface separation $L$ for different sphere radii.
The solid curves from bottom to top correspond
to increasing values of sphere radii. The PFA prediction corresponds to the
 dashed curve. [Colors online].}
\label{Fig_final}
\end{figure}

In order to highlight this striking feature, we plot
 in Fig.~\ref{Fig_final} the ratio between the thermal Casimir force values $F^\plas$ calculated with the plasma model and
$F^\Drud$ obtained with the Drude model, as a function of the distance $L,$ and for different values of the sphere radius $R$ increasing from bottom to top.
The fact that $F^\plas/F^\Drud$ depends on $R$ is a clear signature of the interplay between geometry, temperature and dissipation.
We also show the ratio computed within PFA (dashed line), which approaches 2 at large distances
since the PFA result is proportional to the energy for parallel planes. On the other hand, the exact ratio approaches 3/2 for large radii $R\gg \lambda_P$
and more generally a value between 1 and 3/2 obtained from (\ref{Mieplasma2}) and (\ref{F_Drud}):
\[
\frac{F^\plas}{F^\Drud}\approx \frac{3}{2} \left( 1 + \frac{1}{\alpha^2}-\frac{\coth\alpha}{\alpha} \right) ~ ,~   L \gg \lambda_T, R .
\]
The factor 2 predicted by PFA in the large-distance limit  is never approached by our exact results.

\section{Conclusion}

In this paper, we have used the scattering approach to compute the Casimir free energy and force in the plane-sphere geometry, taking into account both the non zero temperature and the metallic nature of reflectors. 

For the simpler case of perfect metals at intermediate distances, we observe from numerical computations a strong correlation between thermal and geometry effects, and negative entropy values for small spheres, which are clearly not related to dissipation but rather to geometry itself. For small spheres, thermal photons provide a repulsive contribution, thus diminishing the total attractive Casimir force. Those results are endorsed by analytical derivations in the long distance regime.

For the case of Drude metals, evidence of correlations between temperature and dissipation, qualitatively different from those in the parallel plates geometry, is given. As a consequence, the results for the loss less plasma and full Drude models are closer to each other than in the parallel-plate geometry, with the long distance
ratio of 2 reduced to at most 3/2. If these results also hold in the experimental range $R/L> 10^2$, it might diminish the discrepancy between experimental results and predictions of the thermal Casimir force using the Drude model.

\acknowledgments
 The authors thank  I. Cavero-Pelaez and  G.-L. Ingold  for
discussions, CAPES-COFECUB and the French Contract
ANR-06-Nano-062 for financial support, and the ESF Research Networking Programme CASIMIR
(www.casimir-network.com) for providing excellent opportunities for
discussions on the Casimir effect and related topics. P.A.M.N. thanks H.M. Nussenzveig for discussions and  CNPq and Faperj for financial support.

\appendix

\section{Matrix elements of the round-trip operator ${\cal M}^{(m)}$}

In this appendix, we derive explicit expressions for the
matrix elements $\cM^{(m)}(\xi)_{1,2} $ given by (\ref{Mintegral}).
The different coefficients appearing in this equation are first calculated for
real  frequencies $\omega.$
Given values of  $\omega$ and $\bk(k,\varphi)$  define two directions
${\mathbf{\hat K}}(\theta^{\pm},\varphi)$
in reciprocal space
with
\begin{equation}
\label{theta}
\sin\theta^{\pm}=\frac{ck}{\omega},\;\;\;\cos\theta^{\pm}=\pm\frac{ck_z}{\omega},\;\;\; k_z\equiv\sqrt{\omega^2/c^2-k^2} .
\end{equation}

We first consider
 the matrix elements
  implementing the change from the multipole to the plane wave basis.
 The free-space magnetic multipole fields
 have the Fourier representation \cite{Cohen}
($\bK=K {\mathbf{\hat K}}(\theta,\varphi))$)
\begin{eqnarray}
\label{IE}
\mathbf{I}_{\omega\ell m}^{(M)}(\bK)&=& (2\pi)^2 \frac{\delta(K-\omega/c)}{\omega/c}\,
\frac{\mathbf{L}_{\mathbf{\hat K}}Y_{\ell m}(\mathbf{\hat K})}{\sqrt{\ell(\ell+1)}}\\
\mathbf{L}_{\mathbf{\hat K}}& \equiv &  i
\left[
\frac{\hat{\theta}}{\sin\theta}\partial_{\varphi} -
\hat{\varphi}_k\partial_{\theta}
\right] .
\end{eqnarray}
The electric multipole fields are in turn given by
$
\mathbf{I}_{K_0\ell m }^{(E)}(\bK)= -\hK\times \mathbf{I}_{\omega\ell m}^{(M)}(\bK).
$

In order to specify the plane-wave basis, we take the usual  TE and TM polarizations adapted to treat the reflection by the plane at $z=0$:
\begin{eqnarray}
&& \hat{\epsilon}_{\TE}=\mathbf{\hat{z}}\times\mathbf{\hat{k}}=\hat{\varphi} \\
&& \hat{\epsilon}_{\TM}=\hat{\epsilon}_{\TE}\times\hK=\hat{\theta} .
\end{eqnarray}
The matrix element $\langle \bk,-,p|\ell,m,P\rangle$ are then obtained from  the
scalar product $\hat{\epsilon}_{p}\cdot \mathbf{I}_{\omega\ell m}^{(P)}(\bK^{(-)})$
with
$\bK^{(-)}= (\omega/c) {\mathbf{\hat K}}(\theta^{-},\varphi)$ after multiplication by the square root
of the Jacobian $dk_z/d(\omega/c)=\omega/(ck_z):$
\begin{eqnarray}
&&
\label{TE_E}
 \langle \bk,-,\TE|\ell m E\rangle = \frac{2\pi m\sqrt{\omega/c}}{k\sqrt{k_z}} \,
 \frac{Y_{\ell m}(\theta^{-},\varphi)}{\sqrt{\ell(\ell+1)}}  \\
&&\label{TE_M}
\langle \bk,-,\TE|\ell m M\rangle = -
\frac{2\pi i}{\sqrt{k_z\omega/c}} \,\frac{\partial_{\theta}Y_{\ell m}(\theta^{-},\varphi)}{\sqrt{\ell(\ell+1)}}\\
&&
\label{TM_E}
\langle \bk,-,\TM|\ell m E\rangle = \langle \bk,-,\TE|\ell m M\rangle \\
\label{TM_M}
&& \langle \bk,-,\TM|\ell m M\rangle = - \langle \bk,-,\TE|\ell m E\rangle .
\end{eqnarray}

We now derive the Mie
 scattering
 matrix elements $\langle \ell m P | {\cal R}_S(\xi) |\bk,+,p \rangle.$
We
write the electric field in terms of the Debye potentials (scalar fields
satisfying Helmholtz equation) $\Pi^{(E)}(\br)$ and $\Pi^{(M)}(\br)$ for the electric and magnetic multipole components.
In order to have the Debye potentials for a
TE-polarized plane wave propagating along an arbitrary direction  $\hK(\theta,\varphi)$ (amplitude $E_0$),
 we take a rotation with Euler angles $\alpha=\varphi,\beta=\theta,\gamma=0.$ In
terms of the coordinates $x',y',z'$ corresponding to the
rotated  axis, the Debye
potentials have the usual form \cite{Bohren} that corresponds to a plane wave propagating along the $z'$ axis linearly polarized along the $y'$ axis.
To write  in terms of the original
 coordinates, we use  the matrix elements of finite rotations and find ($j_{\ell}(Kr)$ are the spherical Bessel functions \cite{Abramowicz})
\begin{eqnarray} \label{Debye_E}
\Pi^{E}_{\bK,\TE}(r,\Theta,\Phi) &=& \frac{E_0}{2K}\sum_{\ell,m}^{\infty} i^{\ell}\sqrt{\frac{4\pi(2\ell+1)}{\ell(\ell+1)}}
j_{\ell}(Kr)
\\
\nonumber
 &\times &  e^{-i m\varphi} (d^{\ell}_{m,1}(\theta) + d^{\ell}_{m,-1}(\theta)) Y_{\ell m}(\Theta,\Phi) \\
 \label{Debye_M}
 \Pi^{M}_{\bK,\TE}(r,\Theta,\Phi) &=& \frac{E_0}{2iK}\sum_{\ell,m}^{\infty} i^{\ell}\sqrt{\frac{4\pi(2\ell+1)}{\ell(\ell+1)}}
j_{\ell}(Kr)
\\
\nonumber
 &\times &  e^{-i m\varphi} (d^{\ell}_{m,1}(\theta) - d^{\ell}_{m,-1}(\theta)) Y_{\ell m}(\Theta,\Phi) .
\end{eqnarray}

We use the same method to derive the Debye potentials for TM incident polarization:
either we take the polarization
along the $x'$-direction  (instead of the
 $y'$-direction as done in the derivation
for TE polarization)
or else take the third Euler angle to be
$\gamma=-\pi/2$ so that the rotated $Oy'$ axis coincides with $\hat{\epsilon}_{\TM}$ instead of $\hat{\epsilon}_{\TE}.$ This
 amounts to introduce an
additional phase $e^{\mp i\gamma}=\pm i$.
Hence the potentials $\Pi^{E,M}_{\bK,\TM}(r,\Theta,\Phi)$   are obtained from (\ref{Debye_E}) and
(\ref{Debye_M})
 by replacing
$d^{\ell}_{m,\pm 1}(\theta)\rightarrow
(\pm i) d^{\ell}_{m,\pm 1}(\theta)$

Since the scattered
 field propagates outward from the sphere, the corresponding potentials are written in terms of the
 spherical  Hankel functions  $h_{\ell}^{(1)}(Kr)$ \cite{Abramowicz}.
 The Debye potentials for the scattered field
 are then obtained
 by considering the boundary conditions at the surface of the sphere.
In the resulting expression,  each partial-wave term  is multiplied by the corresponding Mie coefficient
 $a_{\ell}$ (electric multipoles) or
 $b_{\ell}$ (magnetic multipoles) \cite{Mazolli03}.
 As expected from spherical symmetry, the Mie coefficients do neither depend on $m$ nor on the direction of incidence.
  They are
 written in terms of the Riccatti-Bessel functions  $\psi_{\ell}(\beta)=\beta j_{\ell}(\beta),$ $\zeta_{\ell}(\beta)=\beta h_{\ell}^{(1)}(\beta)$
 evaluated at  the size parameter $\beta=\omega R/c$ and at $n\beta$ \cite{Bohren}:
 \begin{eqnarray}\label{Mie_coefficients_a}
a_{\ell}(\omega) &=&\frac{n \psi_{\ell}(n\beta)\psi'_{\ell}(\beta)- \psi_{\ell}(\beta)\psi'_{\ell}(n\beta)}
{n \psi_{\ell}(n\beta)\zeta'_{\ell}(\beta)- \zeta_{\ell}(\beta)\psi'_{\ell}(n\beta) }\\
\label{Mie_coefficients_b}
b_{\ell}(\omega) & = &\frac{\psi_{\ell}(n\beta)\psi'_{\ell}(\beta)- n\psi_{\ell}(\beta)\psi'_{\ell}(n\beta)}
{ \psi_{\ell}(n\beta)\zeta'_{\ell}(\beta)- n \zeta_{\ell}(\beta)\psi'_{\ell}(n\beta) } .
\end{eqnarray}

From the Debye potentials for the scattered field,  we find the explicit multipole expansion for
the scattered electric field (in position representation)
$
\langle \br | {\cal R}_S | \bK,\, p \rangle,
$
with $p=\TE,\TM$ representing the incident polarization,
which can also be cast into the formal decomposition
\begin{eqnarray}
\label{ampl2}
\langle \br | {\cal R}_S | \bK,\, p \rangle& =& \sum_{\ell m P} \int_0^{\infty} \frac{d\omega}{2\pi c}
\langle \br | \omega \ell m P \rangle  \\
\nonumber
 & &\times  \langle \omega \ell m P | {\cal R}_S |\bK,\,p \rangle,
\end{eqnarray}
where the vector fields $\langle \br | \omega \ell m P \rangle$ are the inverse Fourier transforms of
$\mathbf{I}_{\omega\ell m P}(\bK)$ [see Eq.~(\ref{IE})].
Since the scattering does not change the frequency, the matrix elements of ${\cal R}_S$ have the general form  ($\bK=\bk+k_z \,\mathbf{\hat z},$ $ k_z>0$)
\begin{equation}
\label{connection}
\langle \omega \ell m P | {\cal R}_S |\bK,\,p \rangle = 2\pi \delta(K-\omega/c)\sqrt{ \frac{ck_z}{\omega}}
\langle  \ell m P | {\cal R}_S |\bk,+,\,p \rangle
\end{equation}
with the square root of the Jacobian $dk_z/d(\omega/c)$ providing once more the connection to our plane-wave basis  $|\bk,\pm,p \rangle_{\omega}$
associated to a given frequency $\omega.$

By comparing the explicit expressions for
 $ \langle \br | {\cal R}_S | \bK,\, p \rangle$
with the formal decomposition
(\ref{ampl2}) and taking  (\ref{connection}) into account,
 we find
\begin{eqnarray}
\label{elements_Rs_E}
\langle \ell m E | {\cal R}_S |  \bk,+,\TE\rangle & = &
\sqrt{
\frac{\pi(2\ell+1)}{k_z\omega/c}
}
\,a_{\ell}(\omega)  \\
\nonumber
&\times& e^{-i m\varphi}
\, (d^{\ell}_{m,1}(\theta^+) + d^{\ell}_{m,-1}(\theta^+))\\
\label{elements_Rs_M}
\langle \ell m M | {\cal R}_S |  \bk,+,\TE\rangle & = &
i\sqrt{
\frac{\pi(2\ell+1)}{k_z\omega/c}
}
\,b_{\ell}(\omega)  \\
\nonumber
&\times& e^{-i m\varphi}
\, (d^{\ell}_{m,1}(\theta^+) - d^{\ell}_{m,-1}(\theta^+)).
\end{eqnarray}
For TM polarization, the matrix elements are obtained from (\ref{elements_Rs_E}) and (\ref{elements_Rs_M})
by substituting $d^{\ell}_{m,\pm 1}(\theta)\rightarrow
(\pm i) d^{\ell}_{m,\pm 1}(\theta).$

The Casimir free energy may be written as an integral over the positive frequency semiaxis (which includes the evanescent sector $0\le\omega<ck$).
Using analyticity properties  of the reflection operators for plane and sphere over the upper complex frequency plane, we transform the integral over
real positive frequencies into an integral over complex frequencies $\omega=i\xi+\eta,$
with $\xi$ running from $+\infty$ to $0$ and $\eta\rightarrow 0^+$ keeping the Matsubara poles $i\xi_n$ outside the closed countour of integration
employed in connection with Cauchy theorem \cite{LambrechtNJP06}.
The resulting expression is given by (\ref{depart}), with the matrix elements $\cM^{(m)}(\xi)_{1,2}$ obtained
by taking $\omega\rightarrow i\xi$ and $k_z\rightarrow i\kappa$
in (\ref{TE_E}) -(\ref{TM_M}) and
 (\ref{elements_Rs_E})-(\ref{elements_Rs_M}) and plugging the results into
 (\ref{Mintegral}).   The final explicit expressions are given in Sec.~III.

\newcommand{\REVIEW}[4]{\textrm{#1} \textbf{#2}, #3, (#4)}
\newcommand{\Review}[1]{\textrm{#1}}
\newcommand{\Volume}[1]{\textbf{#1}}
\newcommand{\Book}[1]{\textit{#1}}
\newcommand{\Eprint}[1]{\textsf{#1}}
\def\etal{\textit{et al}}

\end{document}